\documentclass[%
reprint,
superscriptaddress,
nobibnotes,
amsmath,amssymb,
prl,
aps,
]{revtex4-1}

\usepackage{natbib}
\usepackage[english]{babel}
\usepackage[utf8x]{inputenc}
\usepackage{amsmath}
\usepackage{graphicx}
\usepackage{color}
\usepackage{braket}
\usepackage{sidecap}

\addto\captionsenglish{%
}
\begin{document}

\title{A quantum memory with near-millisecond coherence in circuit QED} 
\author{Matthew Reagor}\email{matthew.reagor@yale.edu}
\author{Wolfgang Pfaff}
\author{Christopher Axline}
\author{Reinier~W. Heeres}
\author{Nissim Ofek}
\author{Katrina Sliwa}
\author{Eric Holland}
\author{Chen Wang}
\author{Jacob Blumoff}
\author{Kevin Chou}
\author{Michael~J. Hatridge}
\author{Luigi Frunzio}
\author{Michel~H. Devoret}
\author{Liang Jiang}
\author{Robert~J. Schoelkopf}
\affiliation{Departments of Applied Physics and Physics, Yale University, New Haven, CT 06520, USA}

\date{\today} 

\begin{abstract}
Significant advances in coherence have made superconducting quantum circuits a viable platform for fault-tolerant quantum computing. To further extend capabilities, highly coherent quantum systems could act as quantum memories for these circuits. A useful quantum memory must be rapidly addressable by qubits, while maintaining superior coherence. We demonstrate a novel superconducting microwave cavity architecture that is highly robust against major sources of loss that are encountered in the engineering of circuit QED systems. The architecture allows for near-millisecond storage of quantum states in a resonator while strong coupling between the resonator and a transmon qubit enables control, encoding, and readout at MHz rates. The observed coherence times constitute an improvement of almost an order of magnitude over those of the best available superconducting qubits. Our design is an ideal platform for studying coherent quantum optics and marks an important step towards hardware-efficient quantum computing with Josephson junction-based quantum circuits.
\end{abstract}

\maketitle

\section{Introduction}

The ongoing quest to build a quantum computer demands sustaining the coherence of quantum states while scaling up the system size. Superconducting quantum circuits have experienced enormous improvements in coherence over the last decade, making them a leading contender for the implementation of practical quantum information processing devices. State-of-the-art Josephson junction qubits reach coherence times, $T_2$, of up to one hundred microseconds \citep{Paik:2011hd,Rigetti:2012en} and can be manipulated in nanoseconds \citep{Motzoi:2009iz}. The large ratio of these timescales allows for high-fidelity gate operations \citep{Chow:2009kf} and places superconducting circuits close to the error threshold required for fault-tolerant quantum computation \citep{Barends:2014fu,Riste:2015dx,Corcoles:2015bg}. However, because the overhead of fault-tolerance scales with the error rates \citep{RevModPhys.87.307}, improving coherence still remains imperative. Furthermore, quantum memories with extended storage times could be useful for operations with finite latency, such as encountered in protocols using measurement and digital feedback \citep{Riste:2012kz,Campagne:2013,Steffen:2013}. It is a continuing challenge to engineer superconducting circuits that enable longer coherence times and, crucially, to understand the limitations on the timescales achievable.

A promising route forward is to supplement superconducting qubits with additional, highly coherent systems for quantum state storage \citep[and references therein]{Kurizki31032015}. Although quantum states have been swapped between superconducting qubits and other systems, no quantum memory times that exceed those of the best qubits themselves have been reported as of yet \cite{Kurizki31032015}. Superconducting microwave cavities, which can reach near-second lifetimes \citep{JP:1970kt}, could be effective quantum memories. Because of their high degree of engineerability, cavities can be coupled to qubits with a large amount of control and precision in what is known as 3D circuit quantum electrodynamics (3D cQED) \citep{Paik:2011hd}. Such qubit-cavity systems have been employed for fast (up to several MHz) generation and manipulation of non-classical photonic resonator states  \citep{2015arXiv150301496H}, and cQED compatible cavities have been shown to yield energy decay times surpassing one millisecond \citep{Reagor:2013kf}. In addition, schemes for hardware-efficient quantum error correction have been proposed for such architectures \citep{Mirrahimi:2014js}. However, realizing long resonator coherence times while also introducing the required coupling to qubits has remained an open challenge.

Here we demonstrate a novel microwave cavity quantum memory that stores quantum states with millisecond relaxation and pure dephasing times.  A superconducting transmon qubit is strongly coupled to the cavity mode, allowing for quantum control of the resonator state on the MHz-scale. We characterize the coherence of the resonator with superpositions of its lowest two Fock states, finding $T_2=0.72\pm0.03\,\text{ms}$. We find that the coherence time of the resonator is somewhat reduced from the value anticipated from measurements on uncoupled resonators of this type. By modifying the decay rate and excited state population of the transmon {\em in situ}, we fully trace that enhancement of decoherence rates back to the qubit-resonator coupling.

\begin{figure*}
\centering
\includegraphics[width=6.75in]{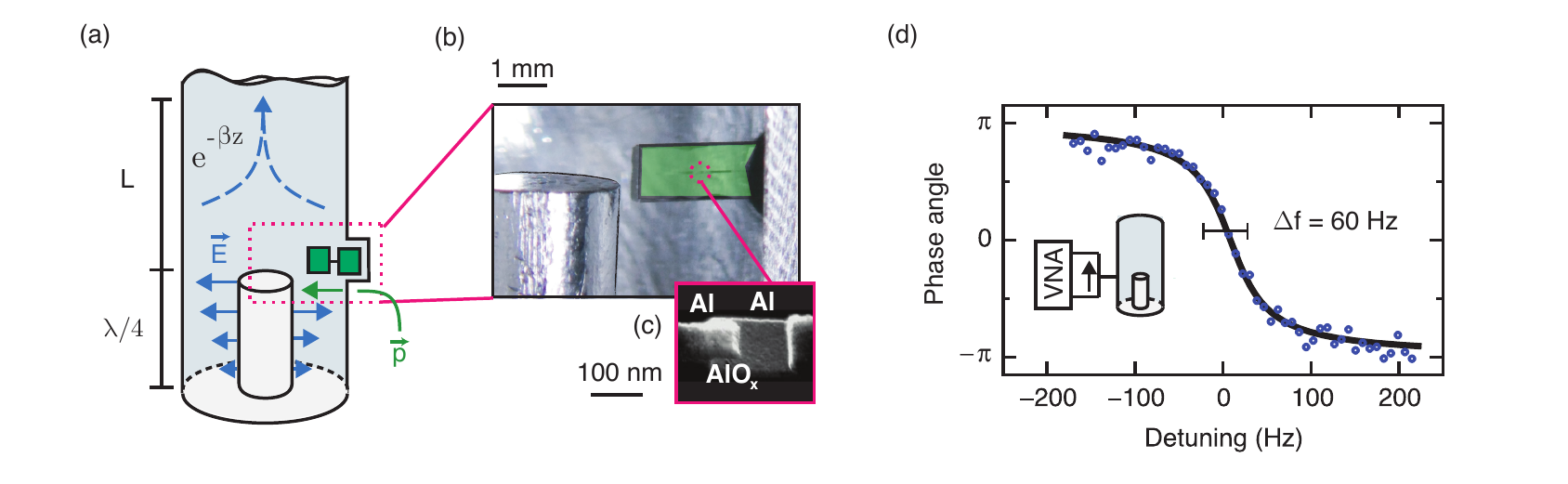}
\caption{System architecture. \textbf{(a)} A quarter-wave coaxial resonator is defined by shorting a coaxial transmission line's inner and outer conductors at one location on the transmission line (bottom) and open-circuiting the line a distance $\lambda/4$ away (upward). A superconducting transmon qubit (green) can be coupled to the $\lambda/4$ mode by aligning the electric dipole moment of the transmon $\protect\overrightarrow{p}$ to the electric field of the resonator $\protect\overrightarrow{E}$. In the section above the coaxial resonator, the outer conductor's cylindrical waveguide TE/TM modes are well below cutoff. Placing a light-tight seam a distance L away from the resonator thus allows the perturbation to be exponentially eliminated. \textbf{(b)} A superconducting transmon qubit on sapphire (green) is inserted through a 1.5 mm hole. The qubit is also coupled to a second cavity used for readout that is not shown here. \textbf{(c)} Electron beam microscopy image of a Josephson junction that provides the nonlinearity to the system. \textbf{(d)}  Measured linear response of a coaxial $\lambda/4$ resonator at single-photon excitation levels, showing a quality factor of $7\times10^{7}$.}
\end{figure*}

\section{High-Q coaxial $\lambda/4$ resonator}

Superconducting microwave cavities can achieve low energy decay rates, $\kappa/2\pi=$ (1-100) Hz, for single microwave photons \citep{Kuhr:2007jm,Reagor:2013kf}. However, leveraging such cavities as quantum memories for circuits has remained an outstanding challenge. Combining resonators with qubits has resulted in resonator performances that are similar to qubits, $\kappa/2\pi\,\gtrsim\,1\,\text{kHz}$ \citep{Kirchmair:2013gu}.  Several dissipation mechanisms are introduced alongside a superconducting qubit, including substrate loss \citep{OConnell:2008jt}, and mechanical instability \citep{Braginsky:1985wy}. Moreover, 3D cQED systems require assembly from parts in order to allow for integration of qubits on chips. Due to finite contact resistance this practice introduces dissipative seams \citep{waveguideflange} that have been identified as a contributor to enhanced decay rates \citep{JP:1970kt,Reagor:2013kf,Brecht:2015}.

In order to avoid this seam dissipation we design our system as sketched in Figure~1. The memory is formed by a coaxial transmission line that is short-circuited on one end and open-circuited on the other by virtue of a narrow circular waveguide \citep{Bianco:1980}. The fundamental resonance frequency, $f_0$, is determined by the length of the transmission line, $\ell\,\approx\,\lambda/4$ (here, $\ell$\,=\,20\,mm results in $f_{0}=4.25\,\text{GHz}$). We rely on a length L of circular waveguide, located between the $\lambda/4$ section and our light-tight seal, to protect the $\lambda/4$ mode from contact resistance at that joint. Because we design the resonator to be well below the waveguide's cutoff frequency ($f_{0} < f_\mathrm c$), the fundamental mode's energy density decreases exponentially into the waveguide section, at a rate determined by the radius of the outer conductor \citep{Pozar:1998uy}, $r=5\,\text{mm}$. We close the device at $L=23\,\text{mm}$ where the normalized energy density of the mode has been reduced to part per billion \citep{SM}. The cavity is fabricated from high purity (4N) aluminum \citep{Reagor:2013kf} and driven by a pin coupler through a hole in the side wall of the cavity \citep{Paik:2011hd}. The mode is strongly under-coupled to avoid external damping ($\kappa_\text{ext}/2\pi \approx 1\,\text{Hz}$). The experiments are performed in a dilution refrigerator at a base temperature of about 15\,mK. Without qubit integration, we determine a single-photon quality factor for one such resonator to be $Q=7\times10^7$ (Fig.~1d) \citep{SM}, corresponding to a single-photon lifetime of approximately 3\,ms. 

\section{Qubit integration}

In addition to the high quality factors achievable, the small mode volume  makes this quarter-wave resonator particularly attractive for integration with transmon qubits. By inserting a sapphire chip holding the qubit as shown in Fig.~1a and Fig.~1b, we are able to achieve strong coupling between the qubit and the memory resonator. The qubit is also coupled to a second, over-coupled cavity used for qubit control and readout \citep{Kirchmair:2013gu}. We set the coupling strength between the transmon and each resonator by the location, orientation, and size of the antenna pads of the transmon. We choose to work in the strong dispersive regime of cQED \citep{Schuster:2007ki}, where the transmon and storage resonator acquire a state-dependent frequency shift that exceeds the line widths of both qubit and resonator, described by the interaction Hamiltonian $\hat{H}_\text{int} / \hbar = -\chi \hat{a}^{\dagger}\hat{a} \hat{b}^{\dagger}\hat{b}$. Here, $\chi$ is the dispersive shift, and $\hat{a}$ and $\hat{b}$ are the photon annihilation operators of the cavity and transmon modes, respectively. For the device discussed in this work, we set the coupling strength to  $\chi/2\pi = 0.5\,\text{MHz}$ (Fig.~2a) \citep{SM}. The resonator can then be controlled on timescales of $\pi/\chi=1\, \mu \mathrm{s}$ \citep{2015arXiv150208015K,2015arXiv150301496H}. This is faster than the coherence time of the transmon, for which we find $T_{2,q}=10~\mu\mathrm{s}$.

\begin{figure}
\includegraphics[width=\columnwidth]{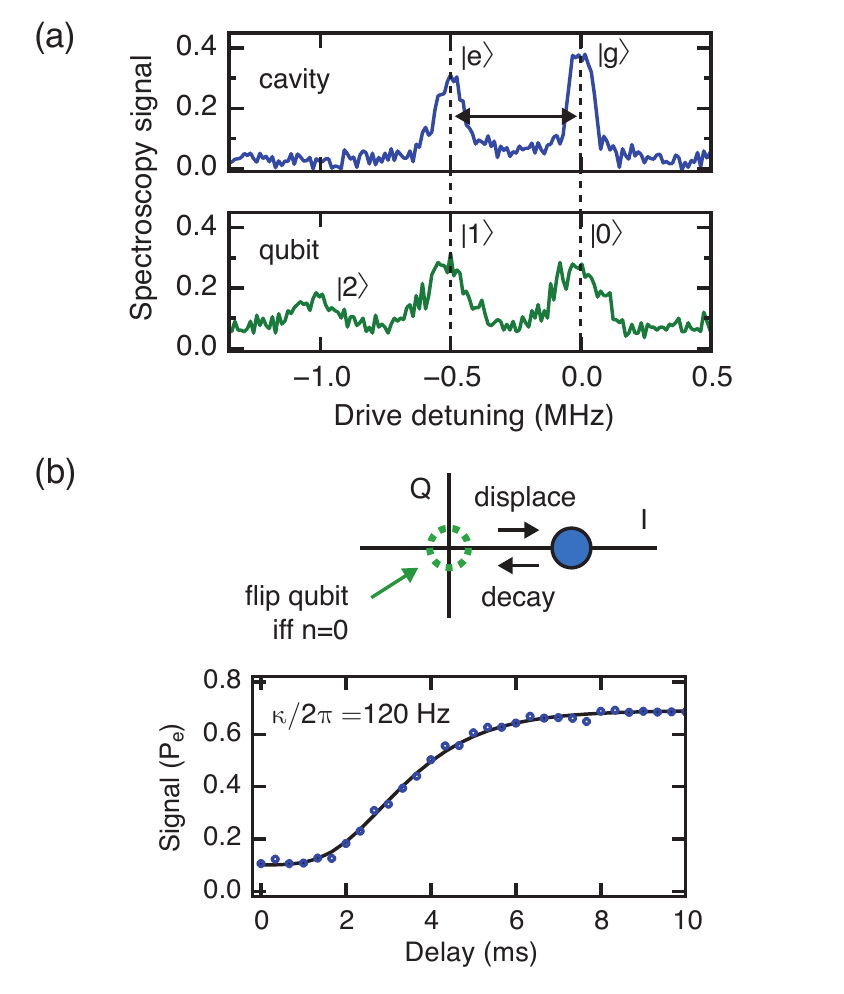}
\caption{Initial characterization of the cavity-qubit system. \textbf{(a)} With strong dispersive coupling, the spectrum of the resonator depends on the state of the transmon (top) and vice versa (bottom). \textbf{(b)} \textbf{(top)} The decay rate of the resonator can be measured by monitoring the population of the vacuum state after displacing the resonator; the state is schematically represented as a blue circle in the IQ plane of the resonator. \textbf{(bottom)} From the population decay we extract a decay rate of $\kappa/2\pi = 120\pm5\,\mathrm{Hz}$, see main text for details.}
\end{figure}

Utilizing the dispersive coupling to the qubit, we can efficiently determine the energy decay rate of the cavity $\kappa$ at the single-photon level (Fig.~2b). We first displace the resonator to create a coherent state with $\beta_{0}=\sqrt{\bar{n}}=3$, where $\beta_0$ is the independently calibrated displacement amplitude \citep{Kirchmair:2013gu} and $\bar n$ is the mean number of photons. This state decays as $\beta(t)=\beta_0 \exp(-\kappa t/2)$, and the probability to find the cavity in its vacuum state is $P_\text{vac}(t) = \exp(-|\beta(t)|^{2})$ \citep{haroche:2006}. We probe the vacuum population directly by applying a photon number-selective $\pi$-pulse on the transmon that is narrow-bandwidth (FWHM $\ll \chi/2\pi$) and centered on the $\ket{g,0}\rightarrow\ket{e,0}$ transition of transmon. This pulse is immediately followed by a measurement of the qubit state \citep{Schuster:2007ki}. The probability for the qubit to be measured in the excited state is $P_\text{e}(t)\propto\exp\left(-|\beta_{0}|^{2} \exp(-\kappa t)\right)$. By fitting the decay curve (Fig.~2b) we extract $\kappa/2\pi=120\pm 5\,\text{Hz}$, corresponding to a quality factor of $Q=3.5\pm0.1\times 10^{7}$. We expect from this classical decay rate that a single excitation in the mode would have a lifetime $T_1=1/\kappa=1.33\pm0.06$\,ms.

\section{Quantum memory characterization}

For evaluating the resonator as a quantum memory it is essential to quantify its coherence time, $T_2$, by the decay of quantum states. The coherence time is bounded by the relaxation time, $T_1 = 1/\kappa$, as $1/T_2 = 1/(2T_1) + 1/T_\phi$.  Analogously to the case of a two-level system, the pure dephasing time (transversal relaxation time), $T_\phi$, is the time constant with which the coherence between a pair of Fock states, $n$ and $n+1$, decays in the absence of  energy relaxation (longitudinal relaxation). $T_1$, $T_2$ can be measured directly by monitoring the time evolution of the resonator after generating the Fock state $\ket 1$ or the superposition state $\left( \ket 0 + \ket 1\right)/\sqrt{2}$, respectively.

We first determine the decay time $T_1$ of the Fock state $\ket{1}$. In our cQED system, arbitrary Fock states (and their superpositions) can be generated in the resonator by a combination of appropriately chosen cavity mode displacements and number-state selective phase gates on the qubit (Fig.~3a) \citep{2015arXiv150208015K,2015arXiv150301496H}. After preparing the state $\ket\psi_0 = \ket 1$ \citep{SM} we measure the time-dependent probability for finding the resonator in the vacuum (Fig.~3b), yielding $T_1 = 1.22\pm0.06\,\text{ms}$, in agreement with the classical energy decay rate extracted from coherent states. The fluxonium superconducting qubit has reached similar $T_1$ at a flux bias of $\Phi=\Phi_0/2$ but has $T_{2,q} \ll T_{1,q}$ at that bias point due to flux noise \citep{Pop:2014}.
 
In order to find the coherence time of the resonator, we prepare the mode in the state $\left( \ket 0 + \ket 1\right)/\sqrt{2}$ and measure the decay of coherence between $\ket 0$ and $\ket 1$ (Fig.~3c) \citep{SM}. This is done in close analogy to Ramsey-type measurements done for qubits \citep{MikeNIke} or resonators \citep{Wang:2013JM}. After state preparation and a delay, we map the phase coherence between $\ket 0$ and $\ket 1$ onto the population of $\ket 0$ with a displacement (taking the role of a $\pi/2$ pulse in a two-level system experiment). This technique gives a detected signal proportional to the state's phase coherence \citep{SM}. The decay of the measured signal yields $T_2 = 0.72\pm0.03\,\text{ms}$ or a pure dephasing time $T_{\phi}=0.98\pm0.05\,\text{ms}$. This coherence time exceeds that of the best superconducting qubit by almost an order of magnitude \citep{Rigetti:2012en}.

\begin{figure}
\includegraphics[width=\columnwidth]{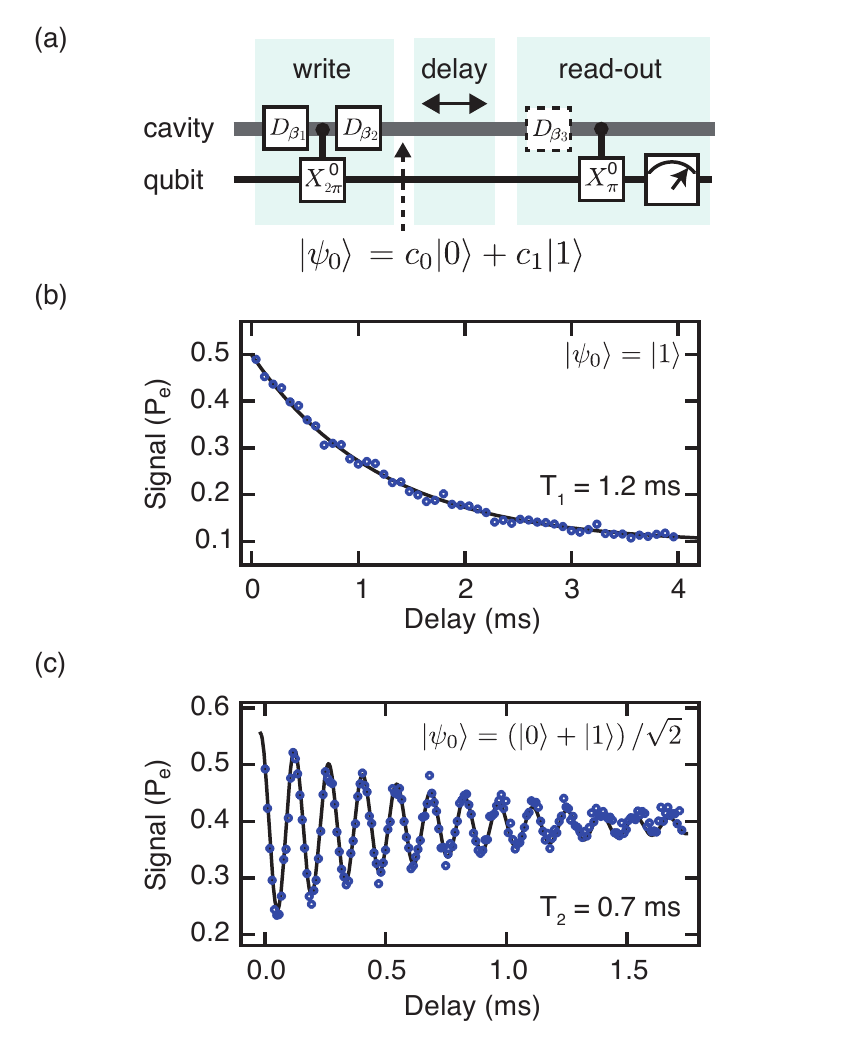}
\caption{Coherence of the quantum memory. \textbf{(a)} Arbitrary Fock state superpositions can be written to the cavity memory using a series of displacements on the cavity and number-selective phase gates on the qubit. \textbf{(b)} To prepare the Fock state $\ket 1$, the optimal choice of displacements is $\beta_{1}=1.14$ and $\beta_{2}=-0.58$ \citep{2015arXiv150208015K,2015arXiv150301496H}. A single-photon decay lifetime of $T_1 = 1.22\pm 0.06\,\text{ms}$ is extracted from the decay of the state. \textbf{(c)} A superposition of the Fock states $\ket 0$ and $\ket 1$ is created with $\beta_{1}=0.56$ and $\beta_{2}=-0.24$. A final displacement of $\beta_{1}=0.8 \exp(j \omega t)$ is used to measure the coherence of the superposition.  This Ramsey-like experiment yields $T_{2} = 0.72\pm0.03\,\mathrm{ms}$, and we infer $T_{\phi} \approx 1\,\mathrm{ms}$.}
\end{figure}

It is critical for our understanding of the cavity memory's performance to identify decoherence mechanisms that are introduced together with the coupling to the qubit. We expect that photon-loss in the resonator should be affected by the dissipation of the transmon as consequence of the hybridization between the modes. This is analogous to the `bad-cavity' limit of cavity QED, where an atom's emission is enhanced by Purcell coupling to a low-Q cavity \citep{Pur46, Goy:1983}. In our case the cavity is longer lived than the artificial atom, leading to an `inverse Purcell effect'. A simple cavity QED treatment \citep{Kimble:1998} would suggest that the resonator mode inherits an additional decay rate from the qubit, $\kappa_\text q \approx (g^2/\Delta^2)\times \gamma$, due to the mode hybridization, where $g$ is the vacuum Rabi rate, $\Delta$ is the detuning, and $\gamma$ the dissipation rate of the qubit.

For a quantitative analysis we perform a numerical finite element simulation of our system, predicting an enhancement of $\kappa_\text q \approx \gamma/600$ \citep{SM}. We can experimentally test this model even without in situ control over $g$ or $\Delta$ (Fig.~4a). Instead, we make use of the higher sensitivity to quasiparticles of the transmon \citep{Lenander:2011dp} to tune $\gamma$ via the base temperature of our refrigerator and observe the resulting change in the decay time of the resonator (Fig.~4a). The linear scaling between the two rates is in quantitative agreement with the prediction from the simulation ($\kappa_{\text{tot}}=\kappa_\text q+\kappa_\text 0$), with the best fit $\kappa_q = (650 \pm 200)^{-1}\gamma$ \citep{SM}. The measurement further yields an intrinsic resonator lifetime of $1/\kappa_0=2\,\text{ms}$, consistent with measurements on empty cavities. These findings suggest that any enhancement in the energy decay rate of the resonator originates from the coupling alone, and is not caused `accidentally' by the methods used to physically assemble the qubit-cavity system.

\begin{figure}
\includegraphics[width=\columnwidth]{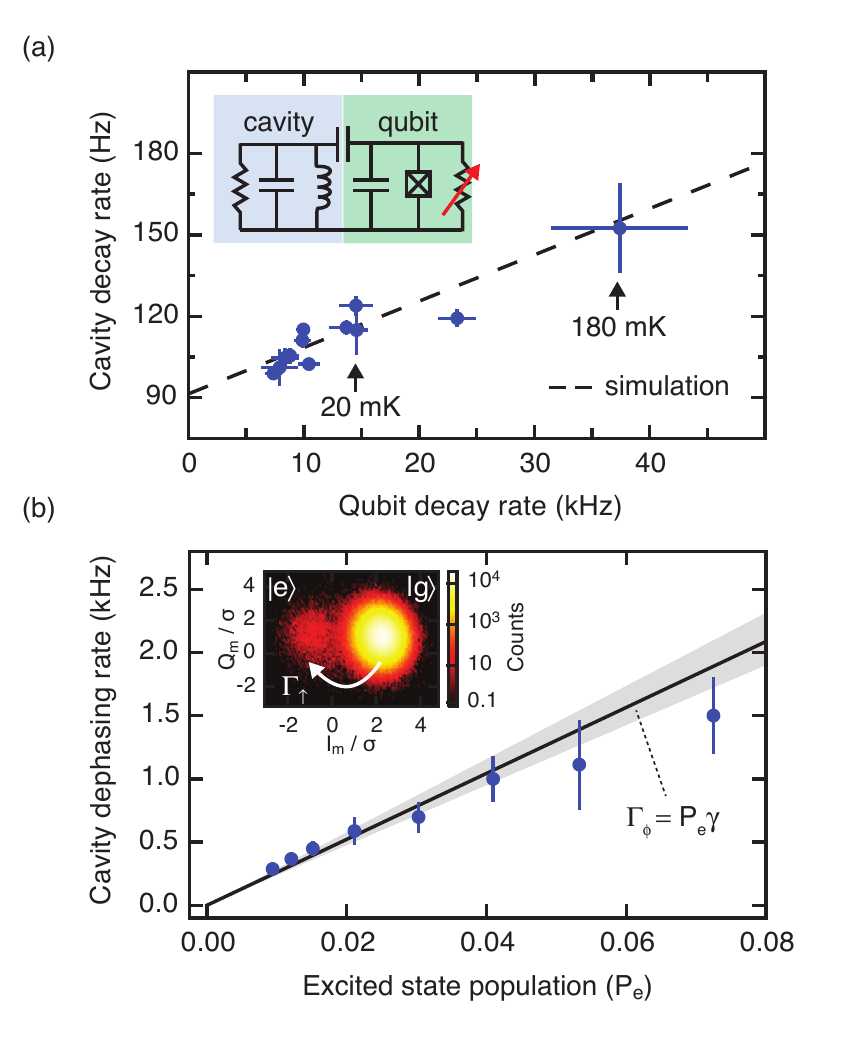}
\caption{\label{fig:4} Qubit-induced decoherence. \textbf{(a) (inset)} We expect the cavity to inherit a photon-loss channel from coupling to the lossy qubit. This effect can be revealed by tuning the decay rate of the qubit {\em in situ}. \textbf{(main)} We measure the decay rates of the cavity and qubit across a temperature range of 20-180~mK. The observed correlation agrees with the predicted qubit-induced loss channel from a three-dimensional electromagnetic simulation of the device \citep{SM}. \textbf{(b) (inset)} A histogram of single-shot measurements of the qubit state, taken in equilibrium with a parametric amplifier, reveals a finite population in the qubit $\ket{e}$ state of $P_{e}=0.8\%$. In the histogram, the measurement records, $\{I_m, Q_m\}$, from the homodyne detection have been scaled by the vacuum fluctuation amplitude, $\sigma$. Quantum jumps due to this thermal bath, $\Gamma_{\uparrow}$, will dephase the state in the cavity memory at rate $P_{e}\kappa_{q}$. \textbf{(main)} The dephasing rate of the cavity memory $\Gamma_{\phi}$ is measured while the qubit is excited weakly via a resonant drive. The observed total cavity dephasing rate, as a function of $P_e$, is in agreement with a model that assumes no intrinsic dephasing in the cavity. For error analysis, see \citep{SM}.}
\end{figure}

Interestingly, we observe a finite pure dephasing time for the resonator state, such that $T_{2}< 2T_{1}$. This means that energy decay is not the only source of decoherence in the resonator, with serious implications for quantum error correction schemes that only assume photon loss \citep{Leghtas:2013ff}. It has been shown that photon shot noise inside a resonator leads to enhanced qubit dephasing in cQED  \citep{Gambetta:2006,Rigetti:2012en,Sears:2012cm}. We investigate whether a thermal excited state population of the transmon \citep{Riste:2012kz,Geerlings:2013kv,Jin:2015hx} is responsible for dephasing  resonator states. This dephasing mechanism can be understood intuitively: if one mode undergoes a stochastic photon number jump, it changes the other mode's frequency by $\chi$, leading to rapid loss of phase information whenever the uncertainty in the time of a stochastic jump is greater than $1/\chi$. The cavity dephasing rate is given by the qubit jump rate as $\Gamma_\phi \approx P_\text e \gamma$, where $P_\text e$ is the excited state population of the qubit \citep{SM}. 

In our sample we estimate an excited state population in equilibrium of $\sim 0.8\%$, corresponding to an effective temperature of 80\,mK (Fig.~4b, inset). In order to test the dependence of $\Gamma_\phi$ on $P_e$, we monitor the $T_2$ of the resonator while applying weak drives on the qubit to populate $\ket{e}$, creating a known increase in qubit jump rate. The total dephasing of the cavity should be given by $\Gamma_\phi=P_\text e \gamma + \Gamma_{\phi}^{0}$, where  $\Gamma_{\phi}^{0}$ is the intrinsic dephasing of the resonator. We find that, to the precision of our measurement, the observed $T_2$ decays are entirely explained by the calibrated $P_\text e$ and observed $\gamma$ (the theory line in Fig.~4b), consistent with the resonator having no intrinsic dephasing mechanisms ($\Gamma_{\phi}^{0}/2\pi\lesssim\,40$\,Hz) \citep{SM}. An encouraging conclusion from these results is therefore that appropriate remedies against thermal qubit population, such as improved thermalization \citep{Barends:2011eh,Corcoles:2011is,Jin:2015hx}, active cooling schemes \citep{Riste:2012kz,Geerlings:2013kv}, or tunable couplers \citep{Mariantoni:2011gu,Srinivasan:2011ik}, could extend the coherence of this quantum memory back to $2T_{1}$, leaving photon loss as the only source of decoherence.

\section{Conclusion and Outlook}

We have demonstrated a long-lived superconducting cavity resonator that can serve as a quantum memory for superconducting quantum circuits, outperforming the best Josephson junction-based qubits available to date. An important conclusion from our data is that any decoherence beyond that of a bare resonator not coupled to a qubit can be explained by the Hamiltonian and qubit properties alone, and is not caused by other technical difficulties that arise from integration. We emphasize that therefore, with given properties of the cavity and the qubit as well as the coupling Hamiltonian, the coherence properties of the memory are optimal, and can only be improved by employing better qubits or cavities. It can be expected that this observation of the coupling-induced decoherence will be useful for ongoing efforts to harness hybrid quantum systems for enhancing the coherence times in superconducting circuits \cite{Kurizki31032015}. 

Because our device reaches the strong dispersive regime of cavity QED, control and measurement can be conducted on fast timescales set by $t=\pi/\chi$. The resonator presented shows little degradation on this timescale, with $\chi T_{1} \approx 3000$, suggesting that quantum operations with very high fidelities can be performed \citep{2015arXiv150301496H}, and that error syndromes on quantum states encoded in microwave photons can be detected much more rapidly than jumps occur \citep{Sun:2014ha}. We therefore expect that the architecture shown will enable further advancement towards fault-tolerance in superconducting quantum computing \citep{Mirrahimi:2014js}, and could enable quantum optics experiments that require very high degrees of control and stability \citep{Raimond:2010hb}.

We thank B. Vlastakis, G. Kirchmair, U. Vool, Z. Leghtas, D. Schuster, and H. Paik for helpful discussions. Facilities use was supported by YINQE and NSF MRSEC DMR 119826. This research was supported by ARO under Grant No. W911NF-14-1-0011. W.P. was supported by NSF grant PHY1309996 and by a fellowship instituted with a Max Planck Research Award from the Alexander von Humboldt Foundation.


%

\pagebreak
\widetext
\begin{center}
\textbf{\large Supplemental materials}
\end{center}
\setcounter{equation}{0}
\setcounter{figure}{0}
\setcounter{table}{0}
\setcounter{page}{1}
\makeatletter
\renewcommand{\theequation}{S\arabic{equation}}
\renewcommand{\thefigure}{S\arabic{figure}}
\renewcommand{\bibnumfmt}[1]{[S#1]}
\renewcommand{\citenumfont}[1]{S#1}


\section{Materials and Methods}

\subsection{Design considerations for waveguide below cutoff}
The resonator geometry described in the main text includes a section of waveguide below cutoff to separate the resonant mode from contact dissipation. The resonator couples to several circular waveguide modes in the section directly above the transmission line (Fig. 1-a in main text). Of these waveguide modes, the TM$_{01}$ mode has the lowest cutoff frequency. Therefore, the TM$_{01}$ mode sets the $\lambda/4$ mode's propagation into the waveguide. The evanescent TM$_{01}$ mode has a propagation constant $\beta=\sqrt{k^{2}-(2.41/a)^2}$, where $k=2\pi/\lambda$ is the wavenumber, and $a=5\,$mm is the radius of the circular waveguide section \citep{Pozar:1998uyS}. At our transmission line's fundamental resonance frequency of 4.25$\,$GHz, the propagation constant is $\beta=\imath/2.03$\,mm, below cutoff. Therefore, the $\lambda/4$ mode's energy density falls as $|E\times H| \propto e^{-2|\beta| z}$ into the waveguide section. Finite element simulations, which take into account all possible waveguide modes, confirm these simple predictions to within $5\%$. We seal the cavity for light-tightness after a length of waveguide section that is $L\approx10/|\beta|$. The resonator's energy has been suppressed at this location by a factor of about $e^{-20}$. We therefore rule out assembly defects such as contact resistance as a potentially limiting dissipation mechanism at internal quality factors of $Q_{\textrm{int}}\sim 10^{9}$.

\subsection{Linear resonator measurements}
In order to estimate the maximum quality of our coaxial transmission line memory, we measure the frequency response of a second, empty resonator with a VNA. We use the shunt-resonator technique to separate the internal quality factor ($Q_{int}$) from the external coupling quality factor ($Q_{ext}$) \citep{Petersan:1998S,Khalil:2012S,Geerlings:2012APLS}. In particular, the data shown in Fig.~1d is evaluated along the frequency-parametrized, complex circle that is traced out in the transmitted linear voltage (S21) in this measurement configuration as done in \citep{Petersan:1998S}. We have achieved $Q_{ext}>10^9$ for this geometry, indicating that the under-coupled control scheme used in the main text is feasible for our control purposes. Indeed, the bandwidth of the curve in Figure~1d in the main text is given by total quality factor of the resonator under test, which is dominated by internal losses $Q_{tot}\approx Q_{int}$. 

In the absence of a transmon or sapphire chip, these coaxial $\lambda/4$ resonators are observed to have single-photon quality factors of $Q_{int}=5-7\times10^{7}$. At large circulating field strengths, corresponding to millions of photons in the resonator, we find that these resonators can have higher performance $Q_{int}=2\times10^{8}$, corresponding to a lifetime of 7\,ms. The saturable nature of this dominate loss mechanism indicates that material defects \citep{Sage:2011S,Wenner:2011S,Megrant:2012S,Bruno:2015S} are a limiting factor for this particular cavity as a memory at the $T_1=Q_{int}/\omega\approx 3\,$ms level. This is consistent with the $T_1=1/\kappa_0=2$\,ms extracted in the main text for the internal loss-limit of the device used in the quantum memory experiment.  

The VNA measurements allow us to place a bound on the quality of the dielectric ($Q_{diel}$) and conductor ($Q_{mag}$) surfaces for our high purity aluminum by assuming that all loss comes from either a dielectric loss ($Q_{int}\propto Q_{diel}$) or conductor loss ($Q_{int}\propto Q_{mag}$). By finite element simulations, we calculate the normalized field-energy stored in the surface layer of our cavity. These calculations allow us to estimate the participation ratios \citep{Gao:2008S,Wenner:2011S} for the AlO$_{x}$ dielectric of $p_{diel}=2\times 10^{-7}$ (assuming an oxide thickness of 3\,nm and a relative dielectric constant of 10), and a kinetic inductance fraction \cite{Day:2003hhS}, or magnetic participation ratio, of $p_{mag}$= 4$\times 10^{-5}$ (using a penetration depth of $50\,$nm \citep{Reagor:2013kfS}). These measurements thus achieve a bound for bulk, high purity aluminum of $Q_{diel}\geq 14$ and $Q_{mag}\geq 3000$ at single-photon energies. Further, the difference between high and low power quality factors indicates that at least one of these loss mechanisms may be saturable, at the $Q_{diel}\sim40$ or $Q_{mag}\sim 8000$ level. 

\subsection{Qubit fabrication}

The transmon qubit is fabricated on a 430\,$\mu$m thick sapphire wafer with a standard Dolan bridge process. A bilayer of resists (MMA/PMMA) support a suspended structure at the Josephson tunnel junction location and are completely stripped where the antenna is to be deposited. Both exposures are completed in a single step of electron-beam lithography. Before deposition, the sapphire surface is cleaned with ion etching, an Ar/O$_2$ descum at 250~V for 30~s. We deposit aluminum with double-angle evaporation ($\pm28^\circ$) with thicknesses of 20~nm and 60~nm, exposing the chamber to oxygen in between these depositions (720 seconds in 2000 Pa static pressure of a gaseous mixture of 85\% argon and 15\% oxygen) and again before removing the sample (600 seconds, 400 Pa). We liftoff aluminum deposited on undeveloped resist. Our transmon's junction has a normal-state resistance of 3.5~k$\Omega$ at room temperature, corresponding to a Josephson inductance of $L_J=4.5$\,nH, or a Josephson energy of $E_J=\Phi_0^{2}/((2\pi)^{2}\hbar L_J)\,=\,150\,\mu eV$, at 15\,mK. The geometry of the transmon's dipole antenna pads are relatively long, approximately 2\,mm each, to achieve strong coupling to both resonators and narrow, 50\,$\mu$m, to maintain a large anharmonicity.   

\subsection{Measurement setup}
Our measurement set up is shown in Supplementary Figure~\ref{fig:setup}. We shape qubit and resonator-control signals with single-sideband modulation (SSB), driven by an FPGA DAC with 500~MHz bandwidth. Stainless steel coaxial cables carry input signals to our package, which is shielded from magnetic field by a light-tight Amuneal magnetic shield (A4K) at the 15~mK stage of the dilution refrigerator. For input lines, we have physical attenuation at the 4~K (20~dB) and 15~mK (30~dB) cooling stages, followed by 10~GHz lowpass filters and eccosorb IR filters. The JPC is pumped by an input line that has physical attenuation at the 4~K stage (20~dB) and the 15~mK stage (23~dB). The pump terminates at the $\Sigma$-port of a hybrid coupler connected to the JPC. 

The qubit state is encoded in the phase of a signal transmitted through the readout cavity. This signal passes through an eccosorb filter, two circulators, and into the $\Delta$ port of of a hybrid connected to the JPC. The JPC acts as a nearly quantum limited phase preserving amplifier \citep{Bergeal:2010S}. It provides the reflected signal with 18~dB gain at 4~MHz bandwidth, which yields sufficient SNR for the single-shot histogram shown in Figure~4b. The readout signal traverses from the JPC back through one circulator, through two additional isolators, and is carried by superconducting coax to a low-nose HEMT amplifier at 4~K. The read-out signal is further amplified at room temperature. Finally, the readout signal is demodulated with a heterodyne interferometer ($f_{\textrm{IF}}$=50~MHz), which is integrated by the FPGA's 1 gigasample-per-second ADC to obtain the histograms shown in the main text. We assign the probability of detecting the qubit in the ground state ($P_{e}$) by thresholding at the center of these two distributions.

\begin{SCfigure} \centering \caption{Experimental schematic. At room temperature, we shape control signals for the qubit and cavity memory by mixing CW local oscillators with a digitized FPGA output, giving full IQ control. The qubit drive is carried on the same input cable as the readout resonator measurement tone. Input signals are filtered at successive dilution refrigerator stages. The transmitted signal is boosted by a parametric amplifier and a following HEMT amplifier, before being demodulated and compared to a reference heterodyne signal in software.} \includegraphics[scale=1.0]{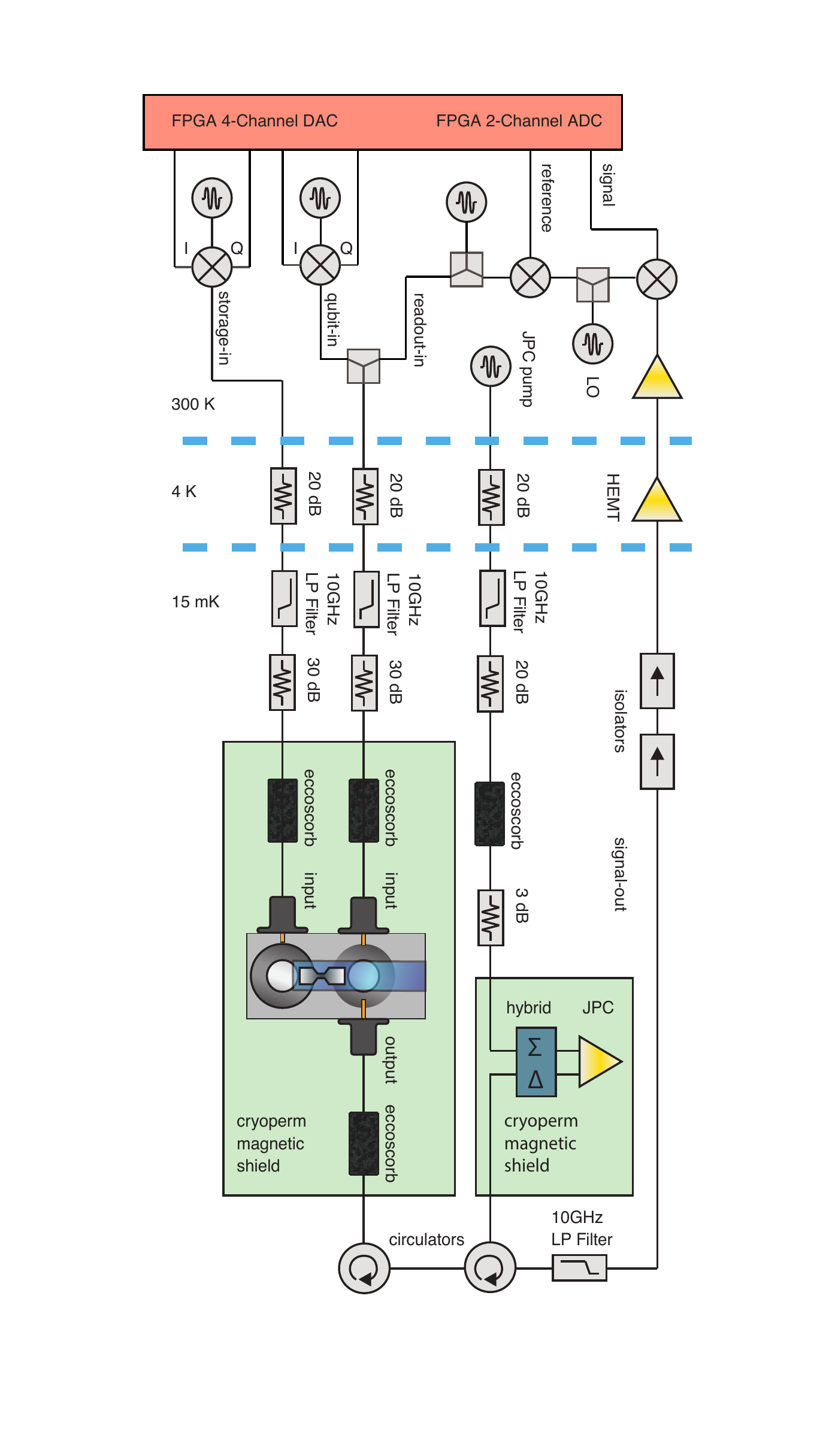} \label{fig:setup}
\end{SCfigure} 

\section{Full system Hamiltonian}

Our system shares the nonlinearity of a single Josepnson junction across the three coupled modes: two nearly-harmonic modes (resonators) and an anharmonic (qubit) mode. Each is well within the weak nonlinearity limit (transmon limit) of the coupled Hamiltonian \citep{Nigg:2012S}, which allows us to approximate the system's total Hamiltonian, as in \citep{Kirchmair:2013guS}, to fourth-order in ladder operators as

\begin{align}\label{eq:S1}\tag{S1}
\frac{H}{\hbar} &=\;  \omega_s a^\dagger a  +\omega_q b^\dagger b +\omega_r c^\dagger c 
\\&- \chi_{sq} a^\dagger a \; b^\dagger b- \chi_{rq} b^\dagger b \; c^\dagger c - \chi_{sr} a^\dagger a \; c^\dagger c \nonumber
\\&- \frac{K_s}{2} a^\dagger a^\dagger a a - \frac{K_q}{2} b^\dagger b^\dagger b b  - \frac{K_r}{2} c^\dagger c^\dagger c c \nonumber
\end{align}
where we label the operators of storage cavity as $a$, transmon as $b$, and the readout mode as $c$. We provide values for the self-Kerr nonlinearities ($K_i$'s) and cross-Kerr nonlinearities ($\chi_{ij}$'s) in Table~I. We allow for a 0.5$\,$mm machining tolerance in the resonators and a junction capacitance of $C_J=2$\,fF in order to correct the linear frequencies in the simulation, increasing the accuracy of the calculated coupling terms.

In the main text, we limit our consideration of the storage cavity to its lowest two levels, and therefore, $K_a$ does not enter into our analysis. However, we measure the mode's nonlinearity to be small ($K_a=450$\,Hz), which is encouraging for error correction schemes, or quantum optics experiments, that access the full Hilbert space of the resonator.

\begin{table}[h]
\caption{Predicted and extracted parameters for the full device device Hamiltonian (Eq.~\ref{eq:S1}).}
\begin{center}
\begin{tabular*}{0.5\textwidth}{@{\extracolsep{\fill}} c c c c}
\hline \hline
$H/\hbar$ & Experiment (Hz) & Simulation (Hz) & Deviation (\%)\\
\hline
$\omega_s/2\pi$ & 4.250 $\times 10^{9}$ & 4.246 $\times 10^{9}$ & $<$ 1\\
$\omega_q/2\pi$ & 7.906 $\times 10^{9}$ & 7.878 $\times 10^{9}$ & $<$ 1\\
$\omega_r/2\pi$ & 9.777 $\times 10^{9}$ & 9.653 $\times 10^{9}$ & 1\\
$\chi_{sq}/2\pi$& 4.99  $\times 10^{5}$ & 5.56 $\times 10^{5}$ & 11 \\
$\chi_{rq}/2\pi$& 8.25  $\times 10^{5}$ & 7.77 $\times 10^{5}$  &6\\
$\chi_{sr}/2\pi$& - & 1.60  $\times 10^{3}$ & - \\
$K_s/2\pi$      & 4.50  $\times 10^{2}$ & 5.20  $\times 10^{2}$ & 16\\
$K_q/2\pi$      & 1.46  $\times 10^{8}$ & 1.41  $\times 10^{8}$ & 3\\
$K_r/2\pi$      &  - & 1.20  $\times 10^{3}$ & -\\

\hline \hline
\end{tabular*}
\end{center}
\label{parameters}
\end{table}

\section{Fock state preparation}
To make a $\ket{1}$ Fock state, we implement the pulse sequence experimentally demonstrated in \citep{2015arXiv150301496HS}. We first displace the resonator to $\beta$\,=\,1.14, via a Gaussian pulse with width $\sigma_t$\,=\,40\,ns. Then, we apply a 2$\pi$ rotation on the transmon that is selective on the qubit's zero-photon dispersive peak, driving the transition $|g,0\rangle\rightarrow|e,0\rangle$. For this pulse, we use a Gaussian tone with $\sigma_t$\,=\,1.5\,$\mu$s so that the frequency bandwidth remains smaller that the dispersive shift of resonator photons ($\sigma_f\,\approx\,100\,\textrm{kHz}\ll\chi/2\pi=500$\,kHz).  A final displacement, $\beta$\,=\,-0.58 finishes the Fock-state creation. 

\begin{figure}
\includegraphics[scale=1.0]{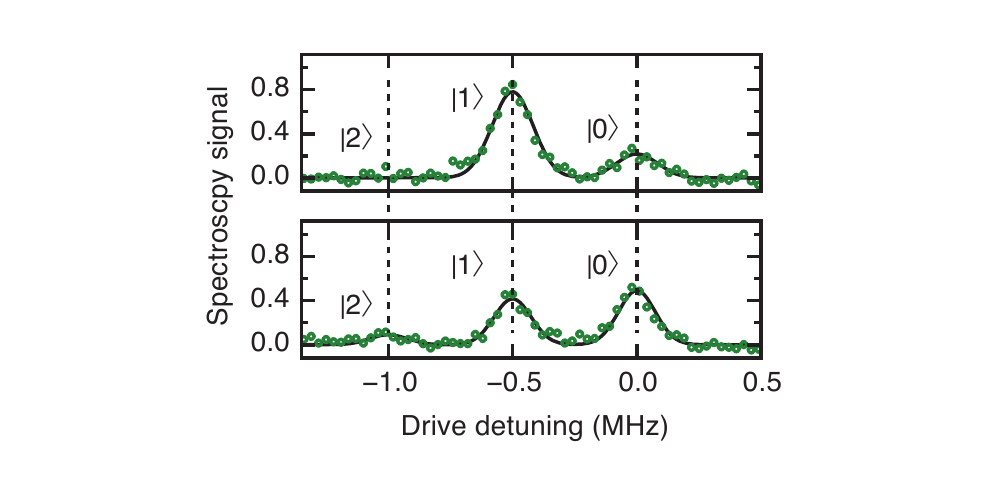}
\caption{\label{fig:fockspec} Preparation of quantum resonator states. To verify the input states for the resonator's $T_1$, $T_2$ experiments, we perform spectroscopy on the qubit after preparing the resonator in an input state.  \textbf{(top)} When the cavity is mostly in the first Fock state, $|1\rangle$, the qubit's frequency distribution reflects these statistics. By the normalized area under each peak, we determine that the population of each Fock state ($P_n$) is $P_0=0.21 \pm 0.02$, $P_1=0.75 \pm 0.02 $, $P_2=0.04 \pm 0.02$ \textbf{(bottom)} After preparing a superposition of $|0\rangle$ and $|1\rangle$, we find the distribution to be $P_0=0.49 \pm 0.02$, $P_1=0.41 \pm 0.02 $, $P_2=0.10 \pm 0.02$. This experiment alone is not sufficient to distinguish between a statistical mixture of these states and a coherent superposition. However, the phase coherence is revealed in the sinusoidal oscillations of the subsequent $T_2$ experiment.}
\end{figure}

In a dissipationless system, this type of sequence can prepare an $\ket{1}$ Fock state with 99$\%$ fidelity \citep{2015arXiv150208015KS}. However, finite qubit coherence ($\chi_{sq}T_{2,q}\,\approx\,$10), limits our quality of state preparation. We quantify the photon-statistics of resulting resonator state by qubit spectroscopy (Sup. Fig.~\ref{fig:fockspec}). 

To make a superposition of $\ket{0}$-$\ket{1}$ Fock states, we modify the pulse sequence above, as outlined in \citep{2015arXiv150208015KS}. We use displacements on the resonator, $\beta$\,=\,0.56,-0.24, interlaced with the same selective 2$\pi$ rotation on the transmon. The statistics of the resulting state is shown in (Sup. Fig.~\ref{fig:fockspec}). We reveal the coherent phase between these two states with a third displacement pulse on the cavity, $\beta=0.8 e^{j \omega t}$, where $\omega$ is a frequency set by a digital phase. The amplitude $|\beta|$ is chosen to saturate the resonator with half of the population in the vacuum when the state has lost all phase coherence.

\section{Dissipation mechanisms}

\subsection{Qubit-induced energy loss}
We estimate the effect of the qubit's dissipation on the resonator by finite element simulations. We use an eigenmode-type solver (HFSS), which is capable of capturing the hybridization between the system's modes and can handle dissipative circuit elements. We add a parallel shunt resistor to the linear inductor that represents the Josephson junction in simulation. The only loss in the calculation arises from currents passing through the inductor. Thus, the ratio of the resulting quality factors is expected to be the scaling between an otherwise perfect cavity's lifetime and the lifetime of the imperfect qubit. 

\begin{figure}
\includegraphics[scale=1.0]{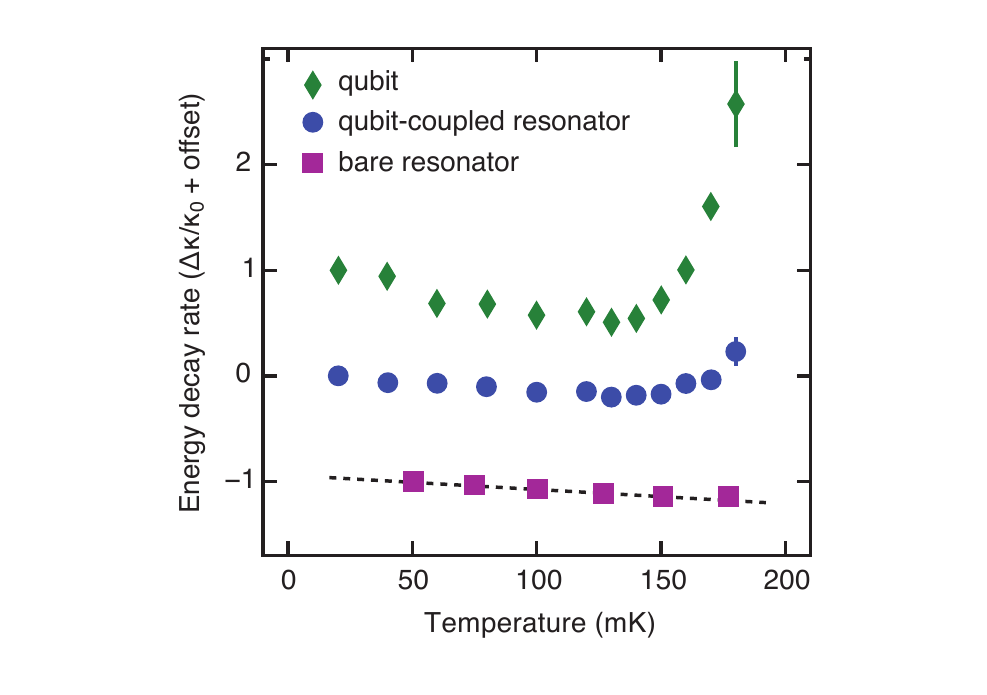}
\caption{\label{fig:tempdep} Temperature dependence of system. We monitor the decay rates of the qubit (green diamonds), qubit-coupled memory resonator (blue circles), and an empty resonator (purple squares) as a function of temperature. We present the relative change in those decay rates, offset to aide visualization. Decreases in the decay rates (improvements in lifetime) are observed for all three modes at elevated temperatures below 150~mK. The dashed line is a best linear fit to the bare resonator's temperature dependence in this range. Above 150\,mK, the quasiparticle sensitivity of the qubit causes at a sharp increase in the decay rate for both the qubit and the qubit-coupled resonator, while the bare resonator continues its trend toward improvement. Figure~4a of the main text is an unnormalized, parameterized version of this plot, with the qubit-coupled resonator decay rates plotted against the qubit decay rates.}
\end{figure}

In the main text, we present temperature dependent data which agrees with these simulations, assuming that the only dissipation in the cavity mode that is temperature dependent is caused by the qubit. However, because the cavity is superconducting, its own BCS temperature dependence might be expected to be observed in this experiment. We have also performed temperature dependence on an empty resonator for comparison (Sup. Fig.~\ref{fig:tempdep}). The linear cavity shows a slight trend toward increased lifetime at elevated temperatures below 180~mK. The bare resonator's lifetime improves up to 15$\%$ at these elevated temperatures. To estimate how this behavior affects the extracted qubit-induced decay rate of the qubit-coupled resonator, we re-express the total decay rate for this coupled resonator as $\kappa_{\textrm{tot}}(T) = \kappa_{\textrm{q}}(T)+ \kappa_{\textrm{0}}(T)$, where the internal resonator dissipation ($\kappa_{\textrm{0}}(T)$) is taken to be the temperature-dependent behavior of an empty resonator. We subtract the best fit linear trend that is obtained from empty resonator measurements, $\kappa_{\textrm{0}}(T)$ (dashed line in Sup. Fig.~\ref{fig:tempdep}), from the observed decay rate of the qubit-coupled resonator, $\kappa_{\textrm{tot}}(T)$. Finally, we correlate the qubit's decay rate to this scaled resonator decay rate for all recorded temperatures. The more thorough analysis changes the best fit result from the main text for $\kappa_{\textrm{q}}=(650\pm200)^{-1}\gamma$ by only $2\%$, well within the $30\%$ uncertainty in the value from our measurements. This effect is therefore ignored in the main text.

\subsection{Qubit-induced dephasing}

An exact model for the pure-dephasing of a qubit due to thermal photons in strongly coupled, lossy resonators has been developed for the dispersive regime of cQED in which we operate\cite{Clerk:2007S,Rigetti:2012enS}. However, because our Hamiltonian is symmetric, this model is directly applicable to resonators being subjected to the reverse process of thermal shot-noise in the qubit mode. The dephasing rate $\Gamma_{\phi}$ derived in \cite{Clerk:2007S,Rigetti:2012enS}, can thus be used for our case of a resonator coupled to a single, thermally populated qubit as
\begin{equation}\tag{S2}
\Gamma_{\phi} = \frac{\gamma}{2} \textrm{Re} \left[\sqrt{\left(1+\frac{2\imath \chi}{\gamma}\right)^{2}+\frac{8\imath \chi P_e}{\gamma}}-1\right],
\end{equation}
where $P_e$ is the excited state population of the qubit and $\gamma$ is the qubit's decay rate. Expanding this expression in the strong dispersive limit ($\chi \gg \gamma$) gives
\begin{equation}\tag{S3}
\Gamma_{\phi} \approx P_e \gamma \left[1-\mathcal{O}\left(\frac{\gamma}{\chi}\right)^{2}\right].
\end{equation}
The quadratic term is of order 1$\times 10^{-4}$ and thus neglected in our analysis. 

\subsection{Error analysis}

The error bounds on all quoted lifetimes in the main text are taken from bootstrapping \citep{Bradley:1994S} five or more experiments of averaged repeated measurements ($\geq$200 averages) under the assumption that the experimental noise is Gaussian. For the qubit's lifetime ($T_{1,q}$) we observe a Gaussian variation ($T_{1,q}\,=\,6.1\pm0.5\,\mu$s) across the 24\,hours of integration required for the transmon dephasing experiment (Fig.~4b in main text). To obtain those statistics, ninety-four $T_{1,q}$ experiments were conducted concurrently with the same number of resonator $T_2$ experiments. The fluctuation in $T_{1,q}$ is represented in the gray shading of Figure~4b in the main text, since this effect generates variation in the slope of $\Gamma_{\phi}$ vs $P_e$.

%

\end{document}